\pdfoutput=1
\documentclass[aps,prd,onecolumn,nofootinbib,preprintnumbers]{revtex4}
\usepackage{latexsym,amssymb, amsmath,color,graphicx}
\usepackage{hyperref}
\newcommand{\tmop}[1]{\ensuremath{\mathrm{#1}}}

\DeclareMathOperator{\lsp}{LSP}
\DeclareMathOperator{\SU}{SU}
\begin{document}
\pagestyle{plain}

\preprint{MCTP-12-04}
\title{Discovering Gluino Events at LHC-8 via Disappearing Chargino Tracks}
\author{Gordon Kane}
\author{Ran Lu}
\author{Bob Zheng}

\affiliation{Michigan Center for Theoretical Physics, University of Michigan, Ann Arbor, MI 48109 USA}

\date{\today}
\begin{abstract}
In this note, we advocate a new method for identifying gluino pair production events at the LHC. The method is motivated by and works for theories with heavy squarks and Wino-like LSPs (with nearly degenerate LSP and chargino). Such theories are well motivated and their gluinos typically have a $\mathcal{O}(50\%)$ branching ratio to charged Winos. Observing the track of a long lived charged Wino produced from gluino decay could give a clear identification of a gluino event.  Charged Wino NLSPs produced in colliders can be long-lived enough to leave a reconstructable high $p_T$ charged track before decaying into a soft pion (or a soft lepton) and the LSP, a signature with low SM background. By supplementing the canonical gluino search strategy with a search for these stiff chargino tracks, our results suggest it will be possible to find gluinos with significantly less luminosity. In addition, we describe a procedure for obtaining a kinematic measurement of the gluino mass using the three momenta of the reconstructed chargino tracks. With measurements of the gluino mass and cross section, it will be possible to determine the gluino spin, and confirm that the excess events are indeed due to a spin 1/2 superpartner. It may also be possible to use these stiff Wino tracks to obtain an approximate measurement of the chargino mass, and therefore the LSP (dark matter) mass.
\end{abstract}
 \maketitle

\section{Introduction}

In recent years it has been increasingly recognized that there is strong motivation for scalar superpartners (squarks, etc) to be very heavy.  Most recently it has been shown that compactified string/M-theories generically have scalars with masses equal to the gravitino mass, which in turn must be larger than about 30 TeV to avoid serious moduli and gravitino cosmological problems \cite{Acharya2007,Acharya2008,Acharya2010}.  On the phenomenological side earlier arguments include the ``focus point" ones \cite{Chan1997,Feng1999a,Feng1999b},`` PeV supersymmetry" \cite{Wells2004}, a flavor motivated approach \cite{Cohen:1996vb}, anomaly mediation motivated heavy scalars \cite{Ibe2006,Asai2007,Kadota2011}, and ``split supersymmetry'' \cite{ArkaniHamed2004,ArkaniHamed2004b}, which argues for turning hierarchy and naturalness problems into virtues and puts scalar masses at superhigh scales ``as high as $10^{13}$ GeV''.

The string/M-theory case necessarily includes the soft-breaking scalar masses ${m_{H_u}}, {m_{H_d}}$ being heavy at the compactification scale.  Then the supersymmetric two-doublet Higgs sector has heavy Higgs bosons plus a light Higgs boson whose mass can be calculated for solutions which have a Higgs mechanism.  This led to the prediction $M_h\approxeq 125$ GeV for $\tan\beta \gtrsim 6$ \cite{gordytalk,Kane2011higgs}, and the result that the observed light Higgs boson should be closely SM-like.  These successful predictions significantly strengthen the motivation for heavy scalars (too heavy to be observed at LHC).  Gauginos especially gluinos should be observable at LHC\cite{Feldman:2010uv,Izaguirre:2010nj,Giudice:2010wb,Alves:2010za,Chen:2010kq,Kane2011top}. Such theories also typically have a wino-like LSP.

For wino-like LSPs the chargino and the LSP are nearly degenerate \cite{Randall:1998uk}, separated only by quantum corrections. Then the chargino dominantly decays to the LSP plus a single soft charged pion (or lepton)\cite{Wells1998}, and travels a mean distance $\sim$10 cm from production, with a significant number of charginos giving a stiff charged massive-particle track that travels tens of centimeters.  This property has recently been emphasized by \cite{Cui}  and by \cite{Moroi2011,Ibe2012} in studying pair production of charginos at the LHC.

In this paper we point out that such long stiff chargino tracks will occur in half or more of all gluino decays, and \emph{provide a new method to detect gluinos} in the theories described above. In such theories it should be possible to detect gluinos with 10-20  $\mathrm{fb}^{-1}$ at LHC-8.  Even in compactified string theories with scalars having masses in the tens of TeV, the gluinos themselves decay quickly in the beam pipe\footnote{The method described in this paper only works for short lived gluinos decaying near the primary vertex, and does not apply to theories with long lived gluinos such as split SUSY.} and this method is applicable.  It is also easy to get an approximate measurement of the gluino mass and cross section, and therefore of the gluino spin\cite{Kane2008}. Measuring the spin provides a crucial check that the produced object is indeed a spin 1/2 partner of the colored gluon, and helps confirm that nature is indeed supersymmetric.

In the following we give detailed numbers to document the above. What can actually be done is very detector dependent, so our results are necessarily approximate and require better analyses which will hopefully be performed with the actual data.  One can obtain an approximate measure of the LSP mass from just the kinematics and variables such as $M_{Eff}$ .  Using information on the chargino tracks and their curvature and possible information on the soft charged pion (or the soft leptons)\footnote{The decay is dominantly to $\lsp + \pi$, but there is a small branching ratio to $\lsp + e,\mu + {\rm neutrino}$} from the chargino decay can sharpen results on the chargino mass and therefore the nearly degenerate LSP mass. In this paper we focus on the well motivated theories, with heavy scalars and wino-LSPs. Purely phenomenologically the method would also work for gluino (or squark) searches in any model with chargino and LSP degenerate within less than 2$m_{\pi}$

\section{Searching for Gluino Events with $\widetilde{W}^{\pm}$ Tracks}

Charged tracks resulting from long lived pair produced charginos can provide an unambiguous SUSY signal in particle colliders\cite{Feng1999,Moroi1999,Ibe2006,Asai2007, Asai2008}. Assuming the Higgsino mass parameter $\mu \gtrsim 1$ TeV as is natural in theories with heavy gravitinos\footnote{For example, it was recently shown in \cite{Acharya2011} and \cite{Feldman:2011ud} that $\mu \sim$ TeV arises naturally in certain M-theory compactifications.}, the charged and neutral Wino masses are essentially degenerate at tree level. The largely model independent mass splitting which results from loop corrections is $\delta m \sim 160$ MeV \cite{Feng1999}. To a good approximation the charged Wino decay width can be given by the two body decay width for $\widetilde{W}^{\pm} \rightarrow \widetilde{W}^0 \pi^{\pm}$\cite{Ibe2006}:
\begin{equation}
\Gamma(\widetilde{W}^{\pm}) = \frac{2 G_F^2}{\pi} \cos^2 \theta_c f_\pi^2 \delta m^3 \sqrt{1 - \frac{m_\pi^2}{\delta m^2}},
\end{equation}
where $\theta_c$ is the Cabbibo and angle and the pion decay width $f_\pi \sim 130$ MeV. For large values of $\mu \gtrsim 1$ TeV and moderate values of $\tan \beta$ the loop contribution dominates the mass splitting, resulting in a charged Wino lifetime $c \tau \sim \mathcal{O}(5)$ cm \cite{Feng1999}\footnote{However if $\mu \sim M_2$, there is an $\mathcal{O}({m_W}^2/\mu^2)$ tree level mass splitting which can potentially become $\mathcal{O}(1)$ GeV, significantly decreasing the lifetime such that $c \tau \ll 1$ cm.}.

A prolific source of detectable charged Wino tracks results from three body decays of pair produced gluinos, $p p \rightarrow \tilde{g} \tilde{g} \rightarrow \widetilde{W}^{\pm}, \widetilde{W}^0 +$ jets. In typical heavy squark and Wino LSP models, the gluino has a $\mathcal{O}(50\%)$  branching ratio to $\widetilde{W}^{\pm}$ final states; these events result in many hard objects, and are easily triggered on \cite{Acharya2009,Kane2011top}. The NLO gluino pair production cross section for the mass range of interest, $600 \hspace{1.5mm}\mathrm{GeV} \lesssim m_{\tilde{g}} \lesssim 1000$ GeV, is $\mathcal{O}(10-1000) \hspace{2mm} \mathrm{fb}$. Increasing the LHC center of mass energy from $\sqrt{s} = 7$ TeV to $\sqrt{s} = 8$ TeV approximately doubles the gluino production cross section (as calculated by PROSPINO\cite{PROSPINO}). Thus with 10 $\mathrm{fb}^{-1}$ of luminosity, we should already expect numerous gluino pair production events at LHC-8; roughly half of these gluinos will produce a charged Wino as a 3-body decay product, since the $\SU(2)$ symmetry relates the branching ratio to $q\bar{q}\widetilde{W}^{0}$, $q'\bar{q}\widetilde{W}^{-}$ and $\bar{q}'q\widetilde{W}^{+}$.

In order to detect charged Winos resulting from gluino decays, $\widetilde{W}^{\pm}$ must live long enough to form a track in the inner detector. Specifically, we focus on the detection capabilities of the ATLAS detector. In the barrel region, the inner detector of the ATLAS detector contains three layers of pixel detector at average radii of 5 cm, 9 cm, 12 cm, then there are four layers of semiconductor tracker (SCT), located respectively at 30 cm, 37 cm, 44 cm and 51 cm away from the beam line. The transition radiation tracker (TRT) is located outside the SCT, 55.4 cm away from the beam line \cite{Aad:2008zzm}. If the charged Wino reaches the third layer of the SCT its track can be reconstructed using the information from the pixel detector and the SCT, and its three-momentum can be determined with good resolution \cite{Asai2008bino}. Perhaps it will be possible to use shorter chargino tracks. The soft pion (or lepton) resulting from the chargino decay would also be a distinctive signal if it could be observed, and would give useful information regarding the kinematics of the event.
\begin{figure}[h]
\begin{center}
\includegraphics[scale=0.45]{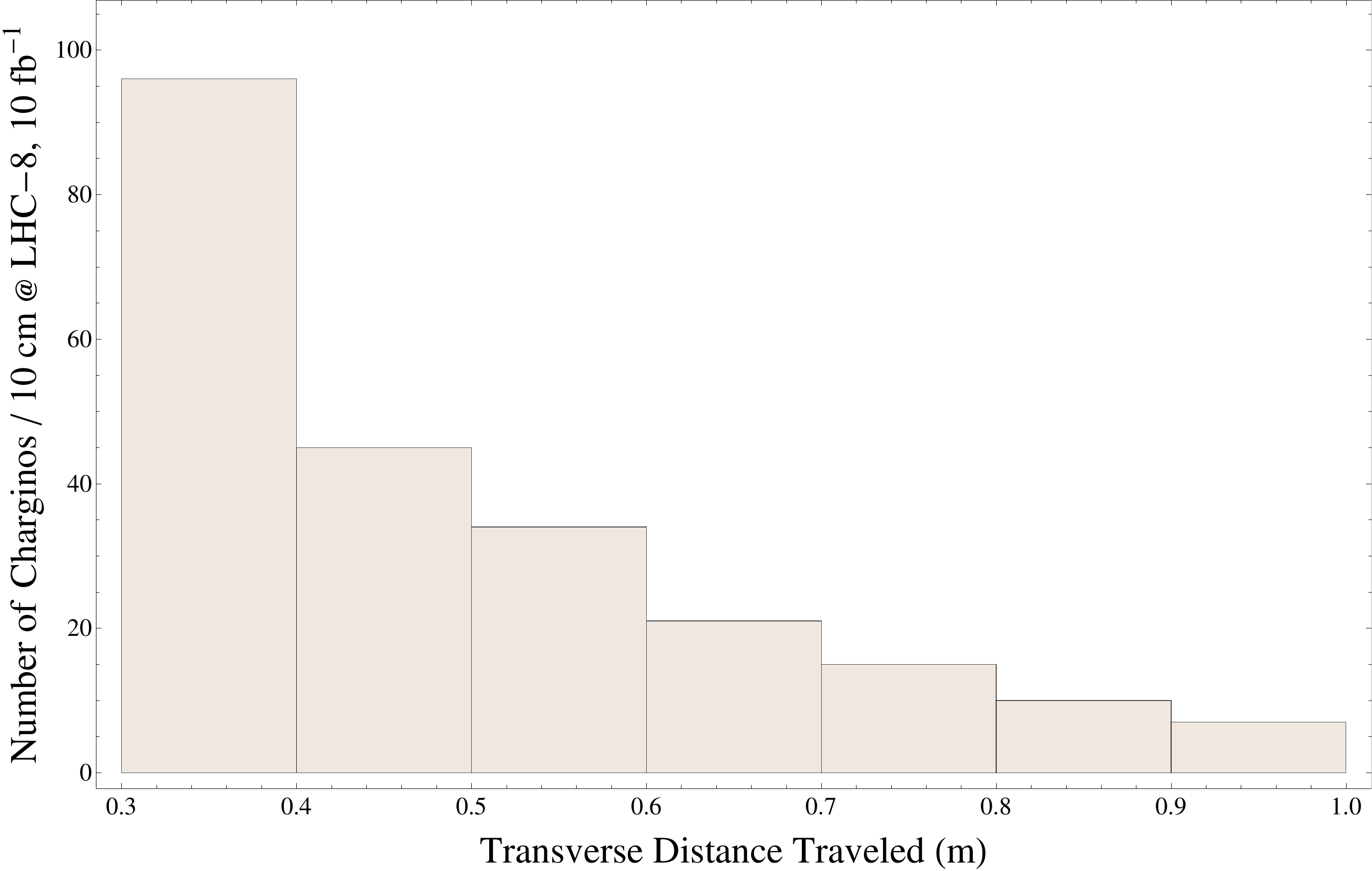}
\end{center}
\caption{Charged Winos resulting from gluino pair production, binned as a function of transverse distance traveled from the beam line. These results correspond to 10 $\mathrm{fb}^{-1}$ of LHC-8 data ($\sigma_{\tilde{g} \tilde{g}} \sim 235$ fb), with $m_{\tilde{g}} = 750$ GeV, $m_{\widetilde{W}} = 150$ GeV. For graphical purposes, charginos traveling a transverse distance $<$ 30 cm are not shown.}
\end{figure}
\begin{table}[h]
\centering
\begin{tabular}{| c | c | c | c | c |}
\hline
$m_{\widetilde{W}}$ &1st SCT Layer & 2nd SCT Layer & 3rd SCT Layer &  4th SCT Layer\\
\hline
100 GeV & 416.3 & 292.6 & 208.2  & 147.9\\
\hline
150 GeV & 232.2 & 150.6 & 98.9 & 69.5\\
\hline
200 GeV & 125.3 & 76.5 & 46.4 & 30.8 \\
\hline
250 GeV & 85.2 & 42.2 & 24.7 & 14.8 \\
\hline
300 GeV & 49.7 &  27.6 & 17.0 & 9.4 \\
\hline
\end{tabular}
\caption{Inclusive count of the number of charginos which make it past a given detector layer. These results correspond to 10 $\mathrm{fb}^{-1}$ of LHC-8 data ($\sigma_{\tilde{g} \tilde{g}} \sim 235$ fb), with $m_{\tilde{g}} = 750$ GeV.}
\end{table}
To estimate the usefulness of looking for stiff charginos, we simulate decays of charged Winos resulting from gluino pair production. The pair produced gluino events are generated by the MadGraph 5/MadEvent package \cite{MadGraph}, and the gluino decays (as well as hadronization and showering) are implemented through PYTHIA6 \cite{Sjostrand:2006za}. The mass splitting $\delta m$ which enters into the decay width is calculated using SOFTSUSY \cite{softsusy}. The results for $m_{\tilde{g}} = 750$ GeV\footnote{We note that with heavy squarks, and the enhancement of gluino decays to heavy quarks due to a lighter stop and sbottom, 750 GeV gluinos are not excluded by any ATLAS or CMS analysis.} with 10 $\mathrm{fb}^{-1}$ of LHC-8 data (without any kinematic cuts) are shown in Figure 1 and Table 1. In Fig. 1, the charged Winos\footnote{In order for the non-thermal cosmology to give approximately correct relic density and not overclose the universe, the chargino cannot be very heavy, so we only consider charginos mass from 100 GeV to 300 GeV\cite{Acharya2008}.} are binned as a function of transverse distance traveled before decay. Table 1 shows an inclusive count of the number charginos which make it past a given detector layer. These results indicate that with the 10 $\mathrm{fb}^{-1}$ of LHC-8 data expected at the end of 2012, for $m_{\widetilde{W}} \lesssim 300$ GeV we expect $\gtrsim 20$ charginos will make it past the third SCT layer.

To search for gluinos, one can use the usual SUSY search channels (or even looser cuts) to isolate relatively pure events from the gluino decay, and then search for chargino tracks in those events. For a concrete example, in the G2-MSSM (resulting from M-theory compactified on a $G_2$ manifold \cite{Acharya2008b}) the gluino has a substantial branching fraction to third generation quarks: $\tilde{g}\rightarrow t\bar{b} W^{-} + h.c.$. As demonstrated in \cite{Acharya2009,Kane2011top}, searching for events containing  a lepton and multiple b-jets is the most powerful way to discover this kind of model. With the additional signals given by chargino tracks, one can also make use of the weaker SUSY search channels such as 1 lepton + 1 b-jet. As an example, we consider a G2-MSSM model with a 750 GeV gluino and 150 GeV Wino LSP with 10 $\mathrm{fb}^{-1}$ of LHC-8 data\footnote{Detector effects are included by running simulated events through the PGS-4\cite{PGS} detector simulation.}.  After applying cuts similar to the 1 lepton + 1 b-jet ATLAS search channel\cite{ATLAS-CONF-2012-003}:
\begin{equation}
1 \hspace{1.5mm}\mathrm{lepton},\hspace{0.5mm} {\not}E_T > 80 \hspace{1.5mm}\mathrm{GeV},\hspace{0.5mm} \ge 4 \hspace{1.5mm}\mathrm{jets},\hspace{0.5mm} \ge 1 \hspace{1.5mm}\mbox{b-jets},
\end{equation}
about 200 gluino pair events survive. Roughly 10 of these surviving events will contain a long chargino track which can be reconstructed. Aside from the 1 lepton + 1 b-jet channel, this chargino signal can be observed in many other SUSY channels, as illustrated in Table 2. In this table, the kinematic cuts defining the signal regions of the 1 b-jet + ${\not}E_T$ channel are defined in \cite{ATLAS-CONF-2012-003}, and the cuts defining the signal regions of the jets + ${\not}E_T$ channel are defined in \cite{ATLAS2011jets}.
\begin{table}[h]
\begin{tabular}{| c | p{2 cm} | p{2 cm} | p{2 cm}|}
\hline
b-jets + ${\not}E_T$ Signal Region\cite{ATLAS-CONF-2012-003} & SR0-A1 \hspace{4mm}$m_{eff}>$ 500 & SR0-B1 \hspace{4mm}$m_{eff}>$ 700& SR0-C1 \hspace{4mm}$m_{eff}>$ 900\\
\hline
Charginos Past 3rd SCT& 28 & 25 & 17 \\
\hline
\end{tabular}
\vspace{4mm}
\begin{tabular}{| c | p{2cm} | p{2cm} | p{2cm} |}
\hline
jets + ${\not}E_T$ Signal Region\cite{ATLAS2011jets} & 4-jet, \hspace{3mm}$m_{eff}>$500 & 4-jet, \hspace{3mm} $m_{eff}>$1000& 4-jet, \hspace{5mm} High Mass\\
\hline
Charginos Past 3rd SCT & 40  & 26 & 20 \\
\hline
\end{tabular}
\caption{Number of charginos which make it past the 3rd SCT layer in signal regions of various ATLAS SUSY search channels. These results are for 10 $\mathrm{fb}^{-1}$ of LHC-8 data. We have chosen channels expected to be sensitive to gluino pair production events. Other search channels give weaker chargino signals, which can be enhanced by loosening cuts. The $m_{eff}$ cuts are given in units of GeV.}
\end{table}
These results demonstrate that the signal of chargino tracks is robust under typical SUSY squark/gluino search cuts; in particular, increasing the $m_{eff}$ cut from 500 GeV to 1000 GeV still maintains $\sim 60\%$ of the chargino signal. Thus chargino tracks provide an additional ``free" channel for gluino discovery, which can be studied without loosening cuts and introducing additional SM background events into the SUSY signal region.

The stub track resulting from chargino decay is a distinctive SUSY signal without SM background. However, in reality the signal may be contaminated by detector noise and mis-reconstructed tracks. We did not include these possible backgrounds in our study; instead we argue they can be controlled without significantly hurting the signals. First of all, the kinematic cuts we applied to remove a large amount of the SM background should also help reduce fake chargino tracks. Requiring $p_T > 80$ GeV for the chargino track candidate keeps $\sim 80 \%$ of the signal events, while eliminating detector noise resulting from soft mis-reconstructed tracks. For a fake track with high $p_T$, the fake chargino $p_T$ will be nearly collinear with ${\not}E_T$ unless there are large contributions to ${\not}E_T$ from energetic neutrinos. Thus imposing a $\Delta\phi \hspace{1mm} (\widetilde{W}^{\pm}, {\not}E_T) > 0.4$ cut will reduce the background from hard jets faking chargino tracks, while keeping $\sim 60 \%$ of the signal events. In addition, the pixel detector can measure $dE/dx$ energy loss of the particle. The charginos on average have $\beta\gamma = 2$, while the SM tracks with the same momentum have much larger $\beta\gamma$. It is possible that this energy loss can be used to separate the real chargino tracks from the backgrounds. With enough events, $dE/dx$ spectra can even be used to estimate the mass of the chargino. However, precise determination of the background after imposing cuts can only be determined by experimental detector groups upon analyzing the data.

\section{Reconstructing the Gluino Mass}
If chargino tracks are observed, one can use this additional information to separate SUSY events from the SM background, and estimate the mass of the gluino even without precise knowledge of the chargino mass. As already mentioned in the ATLAS jets + ${\not}E_T$ search, ignoring the chargino/LSP mass only changes the reconstructed gluino mass by a small amount, $O(m_{LSP}^2/m_{\tilde{g}}^2)$. To obtain sufficient statistics for the analysis, we use a set of cuts slightly looser than the cuts used in the ATLAS jets + ${\not}E_T$ search\cite{ATLAS2011jets}:
\begin{equation}
   {\not}E_T > 130 \tmop{GeV},\; \geq 4 \tmop{jets},\; p_T\left( j1 \right) > 130 \tmop{GeV},\; p_T\left( j2,j3,j4 \right) > 50 \tmop{GeV}, \hspace{1mm} \Delta R(\mathrm{chargino},\hspace{1mm} \mathrm{jets}) > 0.3
    \label{eqn:atlas_jets_met}
\end{equation}
For 10 $\mathrm{fb}^{-1}$ of LHC-8 data, there will be about 60 events containing a reconstructed chargino track which pass the cuts listed in (3). In order to find the pair of jets most likely to  have come from the same gluino as the chargino track, we calculate the transverse momentum of the other chargino/neutralino from the missing transverse momentum, and separate the 6 objects (2 charginos and 4 jets) into 2 clusters, each containing 1 chargino and 2 jets. The optimal clustering is determined by minimizing the difference in $|p_T|$ between the two clusters. Using the cluster containing the real chargino track and assuming $m_{\widetilde{W}} = 0$, we can make a histogram of the reconstructed gluino mass. Fitting the histogram with a distribution gives a reconstructed gluino mass near the actual value, with $\sim100$ GeV uncertainty. Thus chargino tracks can provide an independent kinematic measure of the gluino mass, whose accuracy will improve with improving statistics.

Simply knowing that a gluino of some approximately measured mass decays to a chargino + jets of some $p_T$ puts a valuable upper limit on the chargino mass (and therefore the dark matter mass). With more information, it may be possible to obtain a precise measurement of the chargino mass. For instance, measuring the $p_T$ of the soft pion resulting from the chargino decay along with the chargino track 3-momentum might allow a kinematic reconstruction of the chargino mass. However, it is unclear whether or not measuring such a soft track would be experimentally feasible. Other experimental approaches, such as determining the $dE/dx$ spectra of the chargino tracks, can also refine measurements of the chargino mass.

\section{Concluding Remarks}
We have demonstrated that with the 10-20 $\mathrm{fb}^{-1}$ of LHC-8 data expected by the end of 2012, it is possible to discover gluino pair production events by searching for disappearing chargino tracks from gluino decays which occur in (well-motivated) Wino-like LSP models, in which charginos are nearly degenerate with the LSP. This chargino signal is robust under the typical signal isolation cuts used in SUSY gluino searches, so searching for chargino tracks will enhance canonical SUSY searches for colored superpartners without decreasing signal significance by introducing additional background. Furthermore, measuring the 3-momentum of long lived charginos will give an independent measurement of the gluino mass, and even potentially the chargino/LSP mass. Measuring the gluino mass will allow for determination of the gluino spin once the production cross section is known, providing important verification that the gluino is indeed the supersymmetric partner of the gluon.

\acknowledgments{We would like to thank Aaron Pierce and Junjie Zhu for helpful discussions and advice, and Bobby Acharya, Piyush Kumar and Lian-Tao Wang for comments. The work of G.K., R.L, and B.Z. is supported by the DoE Grant DE-FG-02-95ER40899 and
and by the MCTP. R.L. is also supported by the String Vacuum Project Grant funded
through NSF grant PHY/0917807.}

\bibliography{References}

\begin{thebibliography}{43}
\expandafter\ifx\csname natexlab\endcsname\relax\def\natexlab#1{#1}\fi
\expandafter\ifx\csname bibnamefont\endcsname\relax
  \def\bibnamefont#1{#1}\fi
\expandafter\ifx\csname bibfnamefont\endcsname\relax
  \def\bibfnamefont#1{#1}\fi
\expandafter\ifx\csname citenamefont\endcsname\relax
  \def\citenamefont#1{#1}\fi
\expandafter\ifx\csname url\endcsname\relax
  \def\url#1{\texttt{#1}}\fi
\expandafter\ifx\csname urlprefix\endcsname\relax\def\urlprefix{URL }\fi
\providecommand{\bibinfo}[2]{#2}
\providecommand{\eprint}[2][]{\url{#2}}

\bibitem[{\citenamefont{Acharya et~al.}(2007)\citenamefont{Acharya, Bobkov,
  Kane, Kumar, and Shao}}]{Acharya2007}
\bibinfo{author}{\bibfnamefont{B.~S.} \bibnamefont{Acharya}},
  \bibinfo{author}{\bibfnamefont{K.}~\bibnamefont{Bobkov}},
  \bibinfo{author}{\bibfnamefont{G.~L.} \bibnamefont{Kane}},
  \bibinfo{author}{\bibfnamefont{P.}~\bibnamefont{Kumar}}, \bibnamefont{and}
  \bibinfo{author}{\bibfnamefont{J.}~\bibnamefont{Shao}},
  \bibinfo{journal}{Phys.Rev.} \textbf{\bibinfo{volume}{D76}},
  \bibinfo{pages}{126010} (\bibinfo{year}{2007}), \eprint{hep-th/0701034}.

\bibitem[{\citenamefont{Acharya
  et~al.}(2008{\natexlab{a}})\citenamefont{Acharya, Kumar, Bobkov, Kane, Shao
  et~al.}}]{Acharya2008}
\bibinfo{author}{\bibfnamefont{B.~S.} \bibnamefont{Acharya}},
  \bibinfo{author}{\bibfnamefont{P.}~\bibnamefont{Kumar}},
  \bibinfo{author}{\bibfnamefont{K.}~\bibnamefont{Bobkov}},
  \bibinfo{author}{\bibfnamefont{G.}~\bibnamefont{Kane}},
  \bibinfo{author}{\bibfnamefont{J.}~\bibnamefont{Shao}}, \bibnamefont{et~al.},
  \bibinfo{journal}{JHEP} \textbf{\bibinfo{volume}{0806}}, \bibinfo{pages}{064}
  (\bibinfo{year}{2008}{\natexlab{a}}), \eprint{0804.0863}.

\bibitem[{\citenamefont{Acharya et~al.}(2010)\citenamefont{Acharya, Kane, and
  Kuflik}}]{Acharya2010}
\bibinfo{author}{\bibfnamefont{B.~S.} \bibnamefont{Acharya}},
  \bibinfo{author}{\bibfnamefont{G.}~\bibnamefont{Kane}}, \bibnamefont{and}
  \bibinfo{author}{\bibfnamefont{E.}~\bibnamefont{Kuflik}}
  (\bibinfo{year}{2010}), \eprint{1006.3272}.

\bibitem[{\citenamefont{Chan et~al.}(1998)\citenamefont{Chan, Chattopadhyay,
  and Nath}}]{Chan1997}
\bibinfo{author}{\bibfnamefont{K.~L.} \bibnamefont{Chan}},
  \bibinfo{author}{\bibfnamefont{U.}~\bibnamefont{Chattopadhyay}},
  \bibnamefont{and} \bibinfo{author}{\bibfnamefont{P.}~\bibnamefont{Nath}},
  \bibinfo{journal}{Phys.Rev.} \textbf{\bibinfo{volume}{D58}},
  \bibinfo{pages}{096004} (\bibinfo{year}{1998}), \eprint{hep-ph/9710473}.

\bibitem[{\citenamefont{Feng et~al.}(2000{\natexlab{a}})\citenamefont{Feng,
  Matchev, and Moroi}}]{Feng1999a}
\bibinfo{author}{\bibfnamefont{J.~L.} \bibnamefont{Feng}},
  \bibinfo{author}{\bibfnamefont{K.~T.} \bibnamefont{Matchev}},
  \bibnamefont{and} \bibinfo{author}{\bibfnamefont{T.}~\bibnamefont{Moroi}},
  \bibinfo{journal}{Phys.Rev.Lett.} \textbf{\bibinfo{volume}{84}},
  \bibinfo{pages}{2322} (\bibinfo{year}{2000}{\natexlab{a}}),
  \eprint{hep-ph/9908309}.

\bibitem[{\citenamefont{Feng et~al.}(2000{\natexlab{b}})\citenamefont{Feng,
  Matchev, and Moroi}}]{Feng1999b}
\bibinfo{author}{\bibfnamefont{J.~L.} \bibnamefont{Feng}},
  \bibinfo{author}{\bibfnamefont{K.~T.} \bibnamefont{Matchev}},
  \bibnamefont{and} \bibinfo{author}{\bibfnamefont{T.}~\bibnamefont{Moroi}},
  \bibinfo{journal}{Phys.Rev.} \textbf{\bibinfo{volume}{D61}},
  \bibinfo{pages}{075005} (\bibinfo{year}{2000}{\natexlab{b}}),
  \eprint{hep-ph/9909334}.

\bibitem[{\citenamefont{Wells}(2005)}]{Wells2004}
\bibinfo{author}{\bibfnamefont{J.~D.} \bibnamefont{Wells}},
  \bibinfo{journal}{Phys.Rev.} \textbf{\bibinfo{volume}{D71}},
  \bibinfo{pages}{015013} (\bibinfo{year}{2005}), \eprint{hep-ph/0411041}.

\bibitem[{\citenamefont{Cohen et~al.}(1996)\citenamefont{Cohen, Kaplan, and
  Nelson}}]{Cohen:1996vb}
\bibinfo{author}{\bibfnamefont{A.~G.} \bibnamefont{Cohen}},
  \bibinfo{author}{\bibfnamefont{D.}~\bibnamefont{Kaplan}}, \bibnamefont{and}
  \bibinfo{author}{\bibfnamefont{A.}~\bibnamefont{Nelson}},
  \bibinfo{journal}{Phys.Lett.} \textbf{\bibinfo{volume}{B388}},
  \bibinfo{pages}{588} (\bibinfo{year}{1996}), \eprint{hep-ph/9607394}.

\bibitem[{\citenamefont{Ibe et~al.}(2007)\citenamefont{Ibe, Moroi, and
  Yanagida}}]{Ibe2006}
\bibinfo{author}{\bibfnamefont{M.}~\bibnamefont{Ibe}},
  \bibinfo{author}{\bibfnamefont{T.}~\bibnamefont{Moroi}}, \bibnamefont{and}
  \bibinfo{author}{\bibfnamefont{T.}~\bibnamefont{Yanagida}},
  \bibinfo{journal}{Phys.Lett.} \textbf{\bibinfo{volume}{B644}},
  \bibinfo{pages}{355} (\bibinfo{year}{2007}), \eprint{hep-ph/0610277}.

\bibitem[{\citenamefont{Asai et~al.}(2007)\citenamefont{Asai, Moroi, Nishihara,
  and Yanagida}}]{Asai2007}
\bibinfo{author}{\bibfnamefont{S.}~\bibnamefont{Asai}},
  \bibinfo{author}{\bibfnamefont{T.}~\bibnamefont{Moroi}},
  \bibinfo{author}{\bibfnamefont{K.}~\bibnamefont{Nishihara}},
  \bibnamefont{and} \bibinfo{author}{\bibfnamefont{T.}~\bibnamefont{Yanagida}},
  \bibinfo{journal}{Phys.Lett.} \textbf{\bibinfo{volume}{B653}},
  \bibinfo{pages}{81} (\bibinfo{year}{2007}), \eprint{0705.3086}.

\bibitem[{\citenamefont{Kadota et~al.}(2011)\citenamefont{Kadota, Kane,
  Kersten, and Velasco-Sevilla}}]{Kadota2011}
\bibinfo{author}{\bibfnamefont{K.}~\bibnamefont{Kadota}},
  \bibinfo{author}{\bibfnamefont{G.}~\bibnamefont{Kane}},
  \bibinfo{author}{\bibfnamefont{J.}~\bibnamefont{Kersten}}, \bibnamefont{and}
  \bibinfo{author}{\bibfnamefont{L.}~\bibnamefont{Velasco-Sevilla}}
  (\bibinfo{year}{2011}), \eprint{1107.3105}.

\bibitem[{\citenamefont{Arkani-Hamed and Dimopoulos}(2005)}]{ArkaniHamed2004}
\bibinfo{author}{\bibfnamefont{N.}~\bibnamefont{Arkani-Hamed}}
  \bibnamefont{and}
  \bibinfo{author}{\bibfnamefont{S.}~\bibnamefont{Dimopoulos}},
  \bibinfo{journal}{JHEP} \textbf{\bibinfo{volume}{0506}}, \bibinfo{pages}{073}
  (\bibinfo{year}{2005}), \eprint{hep-th/0405159}.

\bibitem[{\citenamefont{Arkani-Hamed et~al.}(2005)\citenamefont{Arkani-Hamed,
  Dimopoulos, Giudice, and Romanino}}]{ArkaniHamed2004b}
\bibinfo{author}{\bibfnamefont{N.}~\bibnamefont{Arkani-Hamed}},
  \bibinfo{author}{\bibfnamefont{S.}~\bibnamefont{Dimopoulos}},
  \bibinfo{author}{\bibfnamefont{G.}~\bibnamefont{Giudice}}, \bibnamefont{and}
  \bibinfo{author}{\bibfnamefont{A.}~\bibnamefont{Romanino}},
  \bibinfo{journal}{Nucl.Phys.} \textbf{\bibinfo{volume}{B709}},
  \bibinfo{pages}{3} (\bibinfo{year}{2005}), \eprint{hep-ph/0409232}.

\bibitem[{\citenamefont{Kane}(2011)}]{gordytalk}
\bibinfo{author}{\bibfnamefont{G.}~\bibnamefont{Kane}}
  (\bibinfo{publisher}{Presented at the 10th Annual String Phenomenology
  Conference August 22-26, Madison, See
  conferencing.uwex.edu/conferences/stringpheno2011/documents/kane.pdf, slide
  15}, \bibinfo{year}{2011}).

\bibitem[{\citenamefont{Kane et~al.}(2011{\natexlab{a}})\citenamefont{Kane,
  Kumar, Lu, and Zheng}}]{Kane2011higgs}
\bibinfo{author}{\bibfnamefont{G.}~\bibnamefont{Kane}},
  \bibinfo{author}{\bibfnamefont{P.}~\bibnamefont{Kumar}},
  \bibinfo{author}{\bibfnamefont{R.}~\bibnamefont{Lu}}, \bibnamefont{and}
  \bibinfo{author}{\bibfnamefont{B.}~\bibnamefont{Zheng}}
  (\bibinfo{year}{2011}{\natexlab{a}}), \eprint{1112.1059}.

\bibitem[{\citenamefont{Feldman et~al.}(2010)\citenamefont{Feldman, Kane, Lu,
  and Nelson}}]{Feldman:2010uv}
\bibinfo{author}{\bibfnamefont{D.}~\bibnamefont{Feldman}},
  \bibinfo{author}{\bibfnamefont{G.}~\bibnamefont{Kane}},
  \bibinfo{author}{\bibfnamefont{R.}~\bibnamefont{Lu}}, \bibnamefont{and}
  \bibinfo{author}{\bibfnamefont{B.~D.} \bibnamefont{Nelson}},
  \bibinfo{journal}{Phys.Lett.} \textbf{\bibinfo{volume}{B687}},
  \bibinfo{pages}{363} (\bibinfo{year}{2010}), \eprint{1002.2430}.

\bibitem[{\citenamefont{Izaguirre et~al.}(2010)\citenamefont{Izaguirre,
  Manhart, and Wacker}}]{Izaguirre:2010nj}
\bibinfo{author}{\bibfnamefont{E.}~\bibnamefont{Izaguirre}},
  \bibinfo{author}{\bibfnamefont{M.}~\bibnamefont{Manhart}}, \bibnamefont{and}
  \bibinfo{author}{\bibfnamefont{J.~G.} \bibnamefont{Wacker}},
  \bibinfo{journal}{JHEP} \textbf{\bibinfo{volume}{1012}}, \bibinfo{pages}{030}
  (\bibinfo{year}{2010}), \eprint{1003.3886}.

\bibitem[{\citenamefont{Giudice et~al.}(2010)\citenamefont{Giudice, Han, Wang,
  and Wang}}]{Giudice:2010wb}
\bibinfo{author}{\bibfnamefont{G.~F.} \bibnamefont{Giudice}},
  \bibinfo{author}{\bibfnamefont{T.}~\bibnamefont{Han}},
  \bibinfo{author}{\bibfnamefont{K.}~\bibnamefont{Wang}}, \bibnamefont{and}
  \bibinfo{author}{\bibfnamefont{L.-T.} \bibnamefont{Wang}},
  \bibinfo{journal}{Phys.Rev.} \textbf{\bibinfo{volume}{D81}},
  \bibinfo{pages}{115011} (\bibinfo{year}{2010}), \eprint{1004.4902}.

\bibitem[{\citenamefont{Alves et~al.}(2011)\citenamefont{Alves, Izaguirre, and
  Wacker}}]{Alves:2010za}
\bibinfo{author}{\bibfnamefont{D.~S.} \bibnamefont{Alves}},
  \bibinfo{author}{\bibfnamefont{E.}~\bibnamefont{Izaguirre}},
  \bibnamefont{and} \bibinfo{author}{\bibfnamefont{J.~G.}
  \bibnamefont{Wacker}}, \bibinfo{journal}{Phys.Lett.}
  \textbf{\bibinfo{volume}{B702}}, \bibinfo{pages}{64} (\bibinfo{year}{2011}),
  \eprint{1008.0407}.

\bibitem[{\citenamefont{Chen et~al.}(2011)\citenamefont{Chen, Feldman, Liu,
  Nath, and Peim}}]{Chen:2010kq}
\bibinfo{author}{\bibfnamefont{N.}~\bibnamefont{Chen}},
  \bibinfo{author}{\bibfnamefont{D.}~\bibnamefont{Feldman}},
  \bibinfo{author}{\bibfnamefont{Z.}~\bibnamefont{Liu}},
  \bibinfo{author}{\bibfnamefont{P.}~\bibnamefont{Nath}}, \bibnamefont{and}
  \bibinfo{author}{\bibfnamefont{G.}~\bibnamefont{Peim}},
  \bibinfo{journal}{Phys.Rev.} \textbf{\bibinfo{volume}{D83}},
  \bibinfo{pages}{035005} (\bibinfo{year}{2011}), \eprint{1011.1246}.

\bibitem[{\citenamefont{Kane et~al.}(2011{\natexlab{b}})\citenamefont{Kane,
  Kuflik, Lu, and Wang}}]{Kane2011top}
\bibinfo{author}{\bibfnamefont{G.~L.} \bibnamefont{Kane}},
  \bibinfo{author}{\bibfnamefont{E.}~\bibnamefont{Kuflik}},
  \bibinfo{author}{\bibfnamefont{R.}~\bibnamefont{Lu}}, \bibnamefont{and}
  \bibinfo{author}{\bibfnamefont{L.-T.} \bibnamefont{Wang}},
  \bibinfo{journal}{Phys.Rev.} \textbf{\bibinfo{volume}{D84}},
  \bibinfo{pages}{095004} (\bibinfo{year}{2011}{\natexlab{b}}),
  \eprint{1101.1963}.

\bibitem[{\citenamefont{Randall and Sundrum}(1999)}]{Randall:1998uk}
\bibinfo{author}{\bibfnamefont{L.}~\bibnamefont{Randall}} \bibnamefont{and}
  \bibinfo{author}{\bibfnamefont{R.}~\bibnamefont{Sundrum}},
  \bibinfo{journal}{Nucl.Phys.} \textbf{\bibinfo{volume}{B557}},
  \bibinfo{pages}{79} (\bibinfo{year}{1999}), \eprint{hep-th/9810155}.

\bibitem[{\citenamefont{Thomas and Wells}(1998)}]{Wells1998}
\bibinfo{author}{\bibfnamefont{S.~D.} \bibnamefont{Thomas}} \bibnamefont{and}
  \bibinfo{author}{\bibfnamefont{J.~D.} \bibnamefont{Wells}},
  \bibinfo{journal}{Phys.Rev.Lett.} \textbf{\bibinfo{volume}{81}},
  \bibinfo{pages}{34} (\bibinfo{year}{1998}), \eprint{hep-ph/9804359}.

\bibitem[{\citenamefont{Cui et~al.}(2011)\citenamefont{Cui, Randall, and
  Shuve}}]{Cui}
\bibinfo{author}{\bibfnamefont{Y.}~\bibnamefont{Cui}},
  \bibinfo{author}{\bibfnamefont{L.}~\bibnamefont{Randall}}, \bibnamefont{and}
  \bibinfo{author}{\bibfnamefont{B.}~\bibnamefont{Shuve}}
  (\bibinfo{year}{2011}), \eprint{1112.2704}.

\bibitem[{\citenamefont{Moroi and Nakayama}(2011)}]{Moroi2011}
\bibinfo{author}{\bibfnamefont{T.}~\bibnamefont{Moroi}} \bibnamefont{and}
  \bibinfo{author}{\bibfnamefont{K.}~\bibnamefont{Nakayama}}
  (\bibinfo{year}{2011}), \eprint{1112.3123}.

\bibitem[{\citenamefont{Ibe et~al.}(2012)\citenamefont{Ibe, Matsumoto, and
  Yanagida}}]{Ibe2012}
\bibinfo{author}{\bibfnamefont{M.}~\bibnamefont{Ibe}},
  \bibinfo{author}{\bibfnamefont{S.}~\bibnamefont{Matsumoto}},
  \bibnamefont{and} \bibinfo{author}{\bibfnamefont{T.~T.}
  \bibnamefont{Yanagida}} (\bibinfo{year}{2012}), \eprint{1202.2253}.

\bibitem[{\citenamefont{Kane et~al.}(2010)\citenamefont{Kane, Petrov, Shao, and
  Wang}}]{Kane2008}
\bibinfo{author}{\bibfnamefont{G.~L.} \bibnamefont{Kane}},
  \bibinfo{author}{\bibfnamefont{A.~A.} \bibnamefont{Petrov}},
  \bibinfo{author}{\bibfnamefont{J.}~\bibnamefont{Shao}}, \bibnamefont{and}
  \bibinfo{author}{\bibfnamefont{L.-T.} \bibnamefont{Wang}},
  \bibinfo{journal}{J.Phys.G} \textbf{\bibinfo{volume}{G37}},
  \bibinfo{pages}{045004} (\bibinfo{year}{2010}), \eprint{0805.1397}.

\bibitem[{\citenamefont{Feng et~al.}(1999)\citenamefont{Feng, Moroi, Randall,
  Strassler, and Su}}]{Feng1999}
\bibinfo{author}{\bibfnamefont{J.~L.} \bibnamefont{Feng}},
  \bibinfo{author}{\bibfnamefont{T.}~\bibnamefont{Moroi}},
  \bibinfo{author}{\bibfnamefont{L.}~\bibnamefont{Randall}},
  \bibinfo{author}{\bibfnamefont{M.}~\bibnamefont{Strassler}},
  \bibnamefont{and} \bibinfo{author}{\bibfnamefont{S.-f.} \bibnamefont{Su}},
  \bibinfo{journal}{Phys.Rev.Lett.} \textbf{\bibinfo{volume}{83}},
  \bibinfo{pages}{1731} (\bibinfo{year}{1999}), \eprint{hep-ph/9904250}.

\bibitem[{\citenamefont{Moroi and Randall}(2000)}]{Moroi1999}
\bibinfo{author}{\bibfnamefont{T.}~\bibnamefont{Moroi}} \bibnamefont{and}
  \bibinfo{author}{\bibfnamefont{L.}~\bibnamefont{Randall}},
  \bibinfo{journal}{Nucl.Phys.} \textbf{\bibinfo{volume}{B570}},
  \bibinfo{pages}{455} (\bibinfo{year}{2000}), \eprint{hep-ph/9906527}.

\bibitem[{\citenamefont{Asai et~al.}(2008)\citenamefont{Asai, Moroi, and
  Yanagida}}]{Asai2008}
\bibinfo{author}{\bibfnamefont{S.}~\bibnamefont{Asai}},
  \bibinfo{author}{\bibfnamefont{T.}~\bibnamefont{Moroi}}, \bibnamefont{and}
  \bibinfo{author}{\bibfnamefont{T.}~\bibnamefont{Yanagida}},
  \bibinfo{journal}{Phys.Lett.} \textbf{\bibinfo{volume}{B664}},
  \bibinfo{pages}{185} (\bibinfo{year}{2008}), \eprint{0802.3725}.

\bibitem[{\citenamefont{Acharya et~al.}(2011)\citenamefont{Acharya, Kane,
  Kuflik, and Lu}}]{Acharya2011}
\bibinfo{author}{\bibfnamefont{B.~S.} \bibnamefont{Acharya}},
  \bibinfo{author}{\bibfnamefont{G.}~\bibnamefont{Kane}},
  \bibinfo{author}{\bibfnamefont{E.}~\bibnamefont{Kuflik}}, \bibnamefont{and}
  \bibinfo{author}{\bibfnamefont{R.}~\bibnamefont{Lu}}, \bibinfo{journal}{JHEP}
  \textbf{\bibinfo{volume}{1105}}, \bibinfo{pages}{033} (\bibinfo{year}{2011}),
  \eprint{1102.0556}.

\bibitem[{\citenamefont{Feldman et~al.}(2011)\citenamefont{Feldman, Kane,
  Kuflik, and Lu}}]{Feldman:2011ud}
\bibinfo{author}{\bibfnamefont{D.}~\bibnamefont{Feldman}},
  \bibinfo{author}{\bibfnamefont{G.}~\bibnamefont{Kane}},
  \bibinfo{author}{\bibfnamefont{E.}~\bibnamefont{Kuflik}}, \bibnamefont{and}
  \bibinfo{author}{\bibfnamefont{R.}~\bibnamefont{Lu}},
  \bibinfo{journal}{Phys.Lett.} \textbf{\bibinfo{volume}{B704}},
  \bibinfo{pages}{56} (\bibinfo{year}{2011}), \eprint{1105.3765}.

\bibitem[{\citenamefont{Acharya et~al.}(2009)\citenamefont{Acharya, Grajek,
  Kane, Kuflik, Suruliz et~al.}}]{Acharya2009}
\bibinfo{author}{\bibfnamefont{B.~S.} \bibnamefont{Acharya}},
  \bibinfo{author}{\bibfnamefont{P.}~\bibnamefont{Grajek}},
  \bibinfo{author}{\bibfnamefont{G.~L.} \bibnamefont{Kane}},
  \bibinfo{author}{\bibfnamefont{E.}~\bibnamefont{Kuflik}},
  \bibinfo{author}{\bibfnamefont{K.}~\bibnamefont{Suruliz}},
  \bibnamefont{et~al.} (\bibinfo{year}{2009}), \eprint{0901.3367}.

\bibitem[{\citenamefont{Beenakker et~al.}(1996)\citenamefont{Beenakker, Hopker,
  and Spira}}]{PROSPINO}
\bibinfo{author}{\bibfnamefont{W.}~\bibnamefont{Beenakker}},
  \bibinfo{author}{\bibfnamefont{R.}~\bibnamefont{Hopker}}, \bibnamefont{and}
  \bibinfo{author}{\bibfnamefont{M.}~\bibnamefont{Spira}}
  (\bibinfo{year}{1996}), \eprint{hep-ph/9611232}.

\bibitem[{\citenamefont{Aad et~al.}(2008)}]{Aad:2008zzm}
\bibinfo{author}{\bibfnamefont{G.}~\bibnamefont{Aad}} \bibnamefont{et~al.}
  (\bibinfo{collaboration}{ATLAS Collaboration}), \bibinfo{journal}{JINST}
  \textbf{\bibinfo{volume}{3}}, \bibinfo{pages}{S08003} (\bibinfo{year}{2008}).

\bibitem[{\citenamefont{Asai et~al.}(2009)\citenamefont{Asai, Azuma, Jinnouchi,
  Moroi, Shirai et~al.}}]{Asai2008bino}
\bibinfo{author}{\bibfnamefont{S.}~\bibnamefont{Asai}},
  \bibinfo{author}{\bibfnamefont{Y.}~\bibnamefont{Azuma}},
  \bibinfo{author}{\bibfnamefont{O.}~\bibnamefont{Jinnouchi}},
  \bibinfo{author}{\bibfnamefont{T.}~\bibnamefont{Moroi}},
  \bibinfo{author}{\bibfnamefont{S.}~\bibnamefont{Shirai}},
  \bibnamefont{et~al.}, \bibinfo{journal}{Phys.Lett.}
  \textbf{\bibinfo{volume}{B672}}, \bibinfo{pages}{339} (\bibinfo{year}{2009}),
  \eprint{0807.4987}.

\bibitem[{\citenamefont{Alwall et~al.}(2011)\citenamefont{Alwall, Herquet,
  Maltoni, Mattelaer, and Stelzer}}]{MadGraph}
\bibinfo{author}{\bibfnamefont{J.}~\bibnamefont{Alwall}},
  \bibinfo{author}{\bibfnamefont{M.}~\bibnamefont{Herquet}},
  \bibinfo{author}{\bibfnamefont{F.}~\bibnamefont{Maltoni}},
  \bibinfo{author}{\bibfnamefont{O.}~\bibnamefont{Mattelaer}},
  \bibnamefont{and} \bibinfo{author}{\bibfnamefont{T.}~\bibnamefont{Stelzer}},
  \bibinfo{journal}{JHEP} \textbf{\bibinfo{volume}{1106}}, \bibinfo{pages}{128}
  (\bibinfo{year}{2011}), \eprint{1106.0522}.

\bibitem[{\citenamefont{Sjostrand et~al.}(2006)\citenamefont{Sjostrand, Mrenna,
  and Skands}}]{Sjostrand:2006za}
\bibinfo{author}{\bibfnamefont{T.}~\bibnamefont{Sjostrand}},
  \bibinfo{author}{\bibfnamefont{S.}~\bibnamefont{Mrenna}}, \bibnamefont{and}
  \bibinfo{author}{\bibfnamefont{P.~Z.} \bibnamefont{Skands}},
  \bibinfo{journal}{JHEP} \textbf{\bibinfo{volume}{0605}}, \bibinfo{pages}{026}
  (\bibinfo{year}{2006}), \eprint{hep-ph/0603175}.

\bibitem[{\citenamefont{Allanach}(2002)}]{softsusy}
\bibinfo{author}{\bibfnamefont{B.}~\bibnamefont{Allanach}},
  \bibinfo{journal}{Comput.Phys.Commun.} \textbf{\bibinfo{volume}{143}},
  \bibinfo{pages}{305} (\bibinfo{year}{2002}), \eprint{hep-ph/0104145}.

\bibitem[{\citenamefont{Acharya
  et~al.}(2008{\natexlab{b}})\citenamefont{Acharya, Bobkov, Kane, Shao, and
  Kumar}}]{Acharya2008b}
\bibinfo{author}{\bibfnamefont{B.~S.} \bibnamefont{Acharya}},
  \bibinfo{author}{\bibfnamefont{K.}~\bibnamefont{Bobkov}},
  \bibinfo{author}{\bibfnamefont{G.~L.} \bibnamefont{Kane}},
  \bibinfo{author}{\bibfnamefont{J.}~\bibnamefont{Shao}}, \bibnamefont{and}
  \bibinfo{author}{\bibfnamefont{P.}~\bibnamefont{Kumar}},
  \bibinfo{journal}{Phys.Rev.} \textbf{\bibinfo{volume}{D78}},
  \bibinfo{pages}{065038} (\bibinfo{year}{2008}{\natexlab{b}}),
  \eprint{0801.0478}.

\bibitem[{\citenamefont{Conway}()}]{PGS}
\bibinfo{author}{\bibfnamefont{J.}~\bibnamefont{Conway}},
  \emph{\bibinfo{title}{{PGS4}}},
  \urlprefix\url{http://physics.ucdavis.edu/~conway/research/software/pgs/pgs4-general.htm}.

\bibitem[{ATL(2012)}]{ATLAS-CONF-2012-003}
\bibinfo{type}{Tech. Rep.} \bibinfo{number}{ATLAS-CONF-2012-003},
  \bibinfo{institution}{CERN}, \bibinfo{address}{Geneva}
  (\bibinfo{year}{2012}).

\bibitem[{\citenamefont{Aad et~al.}(2011)}]{ATLAS2011jets}
\bibinfo{author}{\bibfnamefont{G.}~\bibnamefont{Aad}} \bibnamefont{et~al.}
  (\bibinfo{collaboration}{ATLAS Collaboration}) (\bibinfo{year}{2011}),
  \eprint{1109.6572}.

\end{thebibliography}
\end{document}